\documentclass[aps,pre,amsmath,amsfonts,amssymb,11pt,nofootinbib,showpacs,letterpaper]{revtex4}
\usepackage{color,epsfig,graphics,psfrag,rotating,theorem}

\def\E{\mathbb E}
\def\prob{{\mathbb P}}

\def\0t{{\tt 0}}
\def\1t{{\tt 1}}
\def\u0t{{\tt \underline{0}}}
\def\H{{\mathbb H}}
\def\ub{\underline{b}}
\def\ux{\underline{x}}
\def\us{\underline{\sigma}}
\def\Z{{\mathbb Z}}

\def\ind{{\mathbb I}}
\def\F{F}

\def\vz{\vec{z}}
\def\vx{\vec{x}}
\def\ot{\leftarrow}
\def\ph{\widehat{\phi}}
\def\reals{{\mathbb R}}
\def\de{{\rm d}}
\def\tL{\widetilde{L}}
\def\ball{{\sf B}}
\def\cball{\overline{\sf B}}
\def\uh{\underline{h}}

\def\A{{\mathbf A}}
\def\B{{\mathbf B}}
\def\b{{\mathbf b}}
\def\bh{\widehat{\mathbf b}}
\def\tB{\widetilde{{\mathbf B}}}
\def\v0{\vec{\tt 0}}
\def\oc{\overline{c}}

\def\F{{\mathbb F}}
\def\J{{\mathbb J}}
\def\D{{\sf D}}
\def\vy{\vec{y}}

\begin{document}

\title{A simple one dimensional glassy Kac model}

\author{Andrea Montanari}
\affiliation{Departments of Electrical Engineering and Statistics,
Stanford University, Stanford CA-9305 USA}

\author{Antoine Sinton}
\affiliation{Laboratoire de Physique Th\'eorique de l'Ecole Normale Sup\'erieure,\\
 24 rue Lhomond 75231 Paris Cedex 05, France}

\begin{abstract}
We define a new family of random spin models with one-dimensional structure,
finite-range multi-spin interactions, and bounded average degree
(number of interactions in which each spin participates). 
Unfrustrated ground states can be described as solutions
of a sparse, band diagonal linear system, thus allowing for 
efficient numerical analysis.

In the limit of infinite interaction range, we recover
the so-called XORSAT (diluted $p$-spin)
model, that is known to undergo a random first order
phase transition as the average degree is increased. 
Here we investigate the most important consequences of a large but finite 
interaction range: $(i)$ Fluctuation-induced corrections to 
thermodynamic quantities; $(ii)$ The need of an inhomogeneous 
(position dependent) order parameter; $(iii)$ The emergence of 
a finite mosaic length scale. In particular, we study the
correlation length divergence at the (mean-field) glass transition.  
\end{abstract}


\maketitle
%
%
\section{Introduction}

A large class of disordered mean field spin models exhibit a behavior
that is reminiscent of the structural glass transition in fragile glasses 
\cite{W1,W2,DynamicsReview}.
As temperature is lowered, they undergo a \emph{`dynamical phase transition'} 
characterized by a diverging relaxation time at a critical
temperature $T_{\rm d}$.
The reason for such a dynamical arrest can in turn be ascribed to
ergodicity breaking: below $T_{\rm d}$  the Boltzmann measure decomposes 
into an exponential number of pure states.
While equilibration is fast within each state, it takes an exponentially large 
(in the system size) time to equilibrate across states.

Below $T_{\rm d}$, the system can be meaningfully
characterized through its \emph{complexity} $\Sigma$,
which gives  the exponential growth rate of the number of pure states
(i.e. the number of such states is about $e^{N\Sigma}$, $N$
being the size). The complexity decreases as temperature is further lowered,
and vanishes linearly at a second (static) transition temperature $T_{\rm s}$. 
This corresponds to an actual thermodynamic phase transition.

A strikingly similar scenario has been found to hold in a large array
of \emph{random constraint satisfaction} 
\cite{BiroliMonassonWeigt,MarcGiorgioRiccardo,OursPNAS} 
problems of interest in 
theoretical computer science\footnote{In a typical constraint 
satisfaction problems, one seeks  an assignment of values to $L$ 
discrete variables in such a way to satisfy $M$ constraints.}. 
The role of temperature is played here 
by the number of constraints per variable $\gamma$, 
while Boltzmann distribution is replaced by the uniform measure over 
solutions of the problems. As the constraint density crosses 
a critical value $\gamma_{\rm d}$, the set of solutions splits into `lumps' 
analogous to pure states. Above a second threshold $\gamma_{\rm s}$
the set of constraints becomes with high probability unsatisfiable.

In the last few years there has been a consistent effort in 
interpreting disordered mean field models as a genuine mean field 
theory for the structural glass transition. This is highly non-trivial since
in any finite-dimensional model there cannot be coexistence of an 
exponentially large number of pure states. 
Imagine trying to select one such state through appropriate
boundary conditions on a box of size $\ell$. This will imply an 
energetic bias towards the selected state, which is of order 
$\sigma\ell^{\theta}$, where $0\le\theta\le d-1$ is a surface tension exponent.
On the other hand, the entropic advantage of the other states is of order
$\Sigma\ell^d$, because of their number. Therefore, for 
$\ell\gtrsim \ell_{\rm s} \equiv (\beta\sigma/\Sigma)^{\frac{1}{d-\theta}}$, 
pure states are no longer stable.

According to the \emph{`mosaic state'} scenario, below $T_{\rm d}$   
a typical configuration of the system can be described as a patchwork
\cite{W3,W4,W5}.
Each patch corresponds to the configuration being close to a particular 
pure state in a localized region whose length scale is $\ell_{\rm s}$. 
Since $\Sigma\sim (T-T_{\rm s})$ at the static transition, the
mosaic lengthscale diverges as  
$\ell_{\rm s} \sim (T-T_{\rm s})^{-\nu}$ with $\nu = 1/(d-\theta)$.

While the mosaic scenario is appealing, its consistency and implications,
as well as its precise meaning, are far from obvious. An important step forward
was achieved in \cite{BB2} where a concrete ``gedanken experiment'' 
was introduced to define $\ell_{\rm s}$. This length scale was interpreted in
\cite{MonSem} as a \emph{point-to-set} correlation length, and its divergence 
was rigorously proved to be equivalent to a 
divergent correlation time. In \cite{FranzMontanari},  $\ell_{\rm s}$ 
was actually shown to diverge at $T_{\rm s}$ in a class of disordered 
Kac models with continuous scalar spins.

Unhappily the models considered in \cite{FranzMontanari} can currently  
be treated only in the Kac limit, and through somewhat 
formal techniques such as the replica method. 
As a consequence, many interesting questions (such as the relevance 
of this limit for realistic interaction ranges, non-perturbative 
fluctuation effects, a precise definition of states) 
cannot be addressed in this context. The present paper aims at introducing 
a new class of models that share some features with the ones
treated in  \cite{FranzMontanari}, while being tractable within 
alternative approaches (e.g. numerically).

We follow the route of generalizing one of the ensembles of random 
constraint satisfaction problems mentioned above, and referred to as 
$k$-XORSAT \cite{XOR_CS,XOR}. We shall require constraints to have finite 
range with respect to an underlying one-dimensional geometry. 
Our motivation is twofold: 
$(i)$ Because of its underlying linear structure, the $k$-XORSAT is 
very well understood. In particular a wealth of informations regarding
pure states and their geometry is accessible through rigorous
techniques \cite{XOR_1,XOR_2,NostroLettera,MontanariSemerjianBethe}; 
$(ii)$ The ensembles of random constraint satisfaction problems 
studied within the computer science and statistical mechanics 
communities have lacked so far  any geometrical structure 
(in physics terms, they are mean field models). 
This is of course a poor cartoon of real world instances, and it
is surely instructive to explore alternative --structured-- models.

Constraint satisfaction problems with finite interaction range
were already considered in \cite{SchwarzMiddleton},
without however considering the interaction range as a parameter.
Further, the most important questions that we shall consider in this paper
were not studied there. 
Several papers \cite{FT1,FT2,FT3} investigated the behavior of
thermodynamic quantities and local order in Kac spin glasses.
Finally a one-dimensional Kac spin glass, with a different (continuous)
phase transition was recently studied numerically in \cite{FranzParisi1d}.

The paper is organized as follows. In Section \ref{sec:Definition}
we introduce our Kac-XORSAT model, and its variants,
and discuss some of their most basic properties in Section \ref{sec:Simple}.
We investigate 
thermodynamic quantities (in particular the ground state entropy)
in Section \ref{sec:GroundState}, and the correlation length divergence in
Section \ref{sec:PointToSet}.
Finally a discussion of our results is presented in \ref{sec:Discussion},
and several technical details are contained in the Appendices.
%
%
\section{Definition of the model}
\label{sec:Definition}

Let us recall that an instance of the $k$-XORSAT problem is defined by 
an $M\times L$ matrix binary $\H$, with row 
weight\footnote{The \emph{row weight} is the number of non-vanishing 
entries in each row of the matrix.} $k$ and a binary vector 
$\ub$ of length $M$.
Solving such an instance requires determining whether the linear system
\begin{eqnarray}
\H\ux = \ub \;\;\;\; \mod\; 2\, ,\label{eq:XORformula}
\end{eqnarray}
admits a solution $\ux\in\{\0t,\1t\}^L$. 
This question is equivalent to asking whether the ground 
state energy of a certain Ising spin model, is zero or not.
More precisely, let $\{i_1(a),\dots,i_k(a)\}\subseteq [L]$ denote the indices 
of the non-zero entries in the $a$-th row of $\H$, 
and $J_a = (-1)^{b_a}$ (here $a\in [M]$, and $[n]\equiv\{1,\dots,n\}$).
The relevant spin model is defined by letting the energy
of configuration $\us \equiv (\sigma_1,\dots,\sigma_L)\in\{+1,-1\}^L$ be
\begin{equation}
E(\us) = \sum_{a=1}^{M} \left( 1 - J_a \sigma_{i_1(a)} \cdots \sigma_{i_k(a)} \right) \, .\label{eq:Energy}
\end{equation}
In the following we shall refer to a particular XORSAT instance as to
a `formula' or a `sample'.

The random $k$-XORSAT (rXOR) 
ensemble is defined by letting $\H$ be a uniformly 
random binary matrix (with dimensions $M\times L$ and row weight $k$)
and $\ub$ a uniformly random vector in $\{\0t,\1t\}^L$. It is also 
useful to consider the \emph{unfrustrated} random $k$-XORSAT
ensemble, defined by setting $\ub=\u0t$ (the all ${\tt 0}$'s vector) 
deterministically. Such an ensemble  exhibits a particularly rich
behavior in the `thermodynamic' limit 
$L\to \infty$, $M\to\infty$ with $\gamma = M/L$ kept fixed.

The \emph{Kac $k$-XORSAT} (KacXOR) ensembles add to the above features
a one-dimensional (or, in linear algebra terms, a band diagonal)
structure. 
One such ensemble is characterized by the 
parameters introduced so far, namely $k$, $L$, and $\gamma$, plus 
an `interaction range' $R$. Unlike in the rXOR ensemble, $\gamma$
is required to be in $[0,1]$, although generalizations are not difficult.
Further, the interaction range is an integer such  that $2R+1\ge k$.
Given such parameters, the matrix $\H$ is sampled as follows.
Rows of $\H$ are indexed by a subset $F$ of $[L]$:
for each $a=1,\dots,L$, $a\in F$ independently from the others with 
probability $\gamma$. In particular, the number of rows of $\H$, $M$,
is a binomial random variable
\begin{eqnarray}
\prob\{M\} = \binom{L}{M}\gamma^M(1-\gamma)^{L-M}\, .
\end{eqnarray}
As $L\to\infty$, the number of rows is 
with high probability\footnote{Following the use in probability theory,
we say that something happens \emph{with high probability} (w.h.p.)
if its probability approaches 1 in the thermodynamic limit.}, 
close to $L\gamma$.
For each $a\in F$, the corresponding row in $\H$ is sampled independently
from the others by letting the indices of non-zero entries
$(i_1(a),\dots,i_k(a))$ be a uniformly random subset of
$\{a-R,\dots,a+R\}$ (i.e. each of the $\binom{2R+1}{k}$ subsets has
the same probability).
We shall refer to $\{a-R,\dots, a+R\}$ as to the \emph{range} of equation $a$.

Finally, 
we let the entries of $\ub$ be indexed by $\F$ as well, and 
iid random in $\{\0t,\1t\}$.  As in the case of random XORSAT,
some simplification is achieved by considering an 
unfrustrated ensemble in which $\ub=\u0t$. 

\begin{figure}
\begin{center}
\setlength{\unitlength}{0.25pt}
\begin{picture}(1500,550)(0,0)
\thicklines
\put(200,150){\circle{50}}
\put(300,150){\circle{50}}
\put(400,150){\circle{50}}
\put(500,150){\circle{50}}
\put(600,150){\circle{50}}
\put(700,150){\circle{50}}
\put(800,150){\circle{50}}
\put(900,150){\circle{50}}
\put(1000,150){\circle{50}}
\put(1100,150){\circle{50}}
\put(1200,150){\circle{50}}
\put(1300,150){\circle{50}}
\thinlines
\put(575,125){\line(0,-1){10}}
\put(575,115){\line(1,0){650}}
\put(1225,125){\line(0,-1){10}}
\put(905,40){\vector(0,1){60}}
\put(850,0){$2 R + 1$}
\put(890,515){$a$}
\put(275,450){\rule{12.5pt}{12.5pt}}
\put(375,450){\rule{12.5pt}{12.5pt}}
\put(475,450){\rule{12.5pt}{12.5pt}}
\put(1175,450){\rule{12.5pt}{12.5pt}}
\put(875,450){\rule{12.5pt}{12.5pt}}
\put(900,450){\line(1,-3){100}}
\put(900,450){\line(-1,-3){100}}
\put(900,450){\line(2,-3){200}}
\put(300,450){\line(2,-3){200}}
\put(300,450){\line(0,-1){300}}
\put(300,450){\line(-1,-3){100}}
\put(400,450){\line(-2,-3){200}}
\put(400,450){\line(2,-3){200}}
\put(400,450){\line(1,-3){100}}
\put(500,450){\line(-1,-3){100}}
\put(500,450){\line(-1,-1){300}}
\put(500,450){\line(1,-1){300}}
\put(1200,450){\line(-2,-3){200}}
\put(1200,450){\line(-1,-1){300}}
\put(1200,450){\line(1,-3){100}}
\put(100,150){\ldots}
\put(100,475){\ldots}
\put(1350,150){\ldots}
\put(1350,475){\ldots}
\end{picture}
\caption{Factor graph representation of a portion
of a KacXOR formula with $k=3$ and $R=3$. Empty circles correspond
to variables (columns of $\H$) and filled squares to equations
(rows of $\H$).}
\label{1ddiag}
\end{center}
\end{figure}
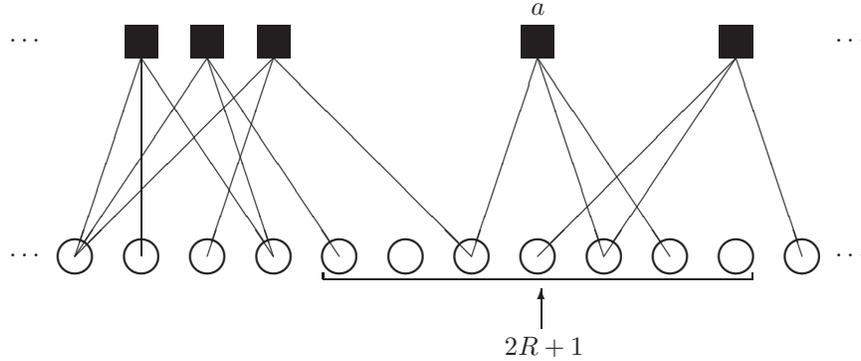
A XORSAT formula admits a natural representation as a factor graph $G$.
This is a bipartite graph including one `parity check node' for each row
in $\H$ (i.e. for each equation in the linear system), and one 
`variable node' for each column (i.e. for each variable in the linear system). 
A parity check and a variable node are connected by an edge if the
corresponding entry of $\H$ is non-vanishing.
An example of such representation is presented in Fig.~\ref{1ddiag}.

There is still one point of the above definition to be clarified.
When $a\le  R$ or $a\ge L-R$, the range for equation $a$
is not included in the sets of variable indices, and it
might be that $i_l(a)\le 0$ or $i_l(a)>L$. We shall consider
two types of boundary conditions. 
With \emph{periodic boundary conditions}, variable indices are interpreted 
modulo $L$. Therefore, if for some index we have $i_l(a)>L$, this
is identified with $i_l(a)-L$, while, if $i_l(a)\le 0$, this 
is identified with $i_l(a)+L$. 

In the case of \emph{fixed boundary condition} we will let
the set potential indices of row of $\H$ be
$\{-R+1,\dots,L+R\}$. Namely $F$ includes each $a$ in this set independently 
with  probability $\gamma$.
To define a fixed boundary condition, we shall 
fix a doubly infinite reference 
configuration\footnote{Notice that the boundary condition  
depends on the reference configuration 
$\ux^{(0)}$ only through $x_{-2R+1}^{(0)},
\dots, x^{(0)}_0$ and  $x_{L+1}^{(0)},
\dots, x^{(0)}_{L+2R}$.} $\ux^{(0)} = \{x_i^{(0)}:
i\in \Z\}$. If in building row $a$ we get an index $i_l(a)\not\in [L]$,
the corresponding non-zero entry is not included in $\H$,
but rather the value $x^{(0)}_{i_l(a)}$ is added to $b_a$.
This corresponds to fixing $\ux=\ux^{(0)}$ `outside' $\{1,\dots,L\}$.
Finally, we shall agree that, whenever considering the 
unfrustrated ensemble, the reference configuration will be
the all $0$'s sequence $\ux^{(0)} = \u0t$.
%
%
\section{Frustrated versus unfrustrated ensemble}
\label{sec:Simple}

The most important feature of the rXOR ensemble in the large size limit
is that it undergoes a SAT-UNSAT phase transition at well defined 
constraint density $\gamma_{\rm s}(k)$. More precisely, a random 
XORSAT formula of the type (\ref{eq:XORformula}) is solvable (SAT) with high 
probability
if $\gamma<\gamma_{\rm s}(k)$, while it is not solvable (UNSAT)
for $\gamma>\gamma_{\rm s}(k)$ \cite{XOR_CS,XOR,XOR_1,XOR_2}.

It is a convenient feature of XORSAT that this phase transition can
be studied by considering uniquely the unfrustrated ensemble.
This simplification can be explained through the well-known identity
\begin{eqnarray}
\prob\{\, \H\,\ux = \ub\, \mbox{ is SAT}\} = \E \left\{2^{L-M}/Z(\H)\right\}\, ,
\label{eq:Harmonic}
\end{eqnarray}
where $Z(\H)$ denotes the number of solutions of the homogeneous linear system
$\H\ux = \u0t$ $\mod 2$. The identity holds irrespective of distribution
of $\H$ provided the right hand side of Eq.~(\ref{eq:XORformula}),
i.e. the vector $\ub$, is uniformly random. In order to
prove it, it is sufficient to notice that  $\H\ux = \ub$ is SAT
if and only if $\ub$ is in the image of $\H$. Since the dimension
of the image of $\H$ is $\text{rank}(\H) = L-\text{dim}\, \text{ker}(\H)$,
this happens with probability  $2^{L-\text{dim}\, \text{ker}(\H)}/2^M$.
on the other hand, $Z(\H) = 2^{\text{dim}\, \text{ker}(\H)}$. 
Equation (\ref{eq:Harmonic}) follows by taking expectation with respect to 
$\H$.

Within the rXOR ensemble, for $\gamma<\gamma_{\rm s}(k)$, $Z(\H)$ 
is tightly concentrated around $2^{L-M}$, implying 
$\prob\{ \H\,\ux = \ub\, \mbox{ is SAT}\}\approx 1$. 
Viceversa for $\gamma>\gamma_{\rm s}(k)$, typically 
$Z(\H) \doteq 2^{L\phi(\gamma)}$ (here $\doteq$ denotes equality
to leading exponential order), with $\phi(\gamma)>1-\gamma$ and therefore
the formula is SAT with exponentially small probability.

Furthermore, as long as the non-homogeneous solution has at least
one solution, its number of solution is independent of $\ub$, and is given by 
$Z(\H)$. Even more, the set of solutions is an affine space obtained by 
translating the linear space of solutions of the homogeneous system.
In other words, conditional to the problem being solvable
(which happens with high probability for $\gamma<\gamma_{\rm s}(k)$)
the frustrated and unfrustrated ensemble are essentially equivalent.

An important novelty within the KacXOR ensemble is that
the linear system (\ref{eq:XORformula}) is always UNSAT 
with high probability if we let $L\to\infty$ with 
$\gamma, R$ fixed. More precisely, we expect that 
\begin{eqnarray}
\prob\{\H\,\ux = \ub\, \mbox{ is SAT}\} \doteq e^{-L\, \Lambda(\gamma,R)}\, ,
\label{eq:SatProbability}
\end{eqnarray}
for some $\Lambda(\gamma,R)$ strictly positive and non-decreasing in 
$\gamma$. This phenomenon was already oserved in \cite{SchwarzMiddleton}
for a related model. The basic reason for this behavior
is that small subsets of nearby rows of $\H$ have a fair chance of being 
linearly dependent. If this is the case, the corresponding linear subsystem
is unsolvable with finite probability. In the large $L$ limit, the expected 
number of such substructures is of order $L$, and the probability that none
is present is exponentially small, thus leading to the above behavior.

It is not difficult to prove the above statement, 
and indeed to prove lower bounds on the rate $\Lambda(\gamma,R)$
by combinatorial techniques. The basic idea is to select a particular type of
substructure that leads to unsatisfiability and estimate the 
probability that no such substructure is present.
The simple such substructure is obtained when two lines of $\H$
coincide but the corresponding entries of $\ub$ do not.
Using Janson inequality this yields
\begin{eqnarray}
\Lambda(\gamma,R) & \ge & K_1\gamma^2-K_2\gamma^3 (1-K_0\gamma^2)^{-1}\, ,
\label{eq:LowerBoundProb}
\end{eqnarray}
where
\begin{eqnarray}
K_0& \equiv &{2R+1-1 \choose k}{2R+1 \choose k}^{-2}\, ,\;\;\;\;\;\;\;
K_1  \equiv  \frac{ 1}{2{2R+1 \choose k}^2} \sum_{p=1}^{2R+1-k} {2R+1-p \choose k}\, ,\label{eq:K0K1}\\
K_2 & \equiv & \frac{3\gamma^3}{8{2R+1 \choose k}^3} \sum_{p=2}^{2R +1 -k} (p-1) {2R+1-p \choose k} \, .\label{eq:K2}
\end{eqnarray}
Such a lower bound  is 
easily seen to be strictly positive for $\gamma$ small enough.
We refer to Appendix \ref{app:Janson} for a derivation of this formula.

Notice that the lower bound in Eq.~(\ref{eq:LowerBoundProb})
vanishes as $1/R^k$ when $R\to\infty$. We expect the same behavior to 
hold for the actual exponent as long as $\gamma$ is below the
(mean-field) satisfiability threshold $\gamma_{\rm s}(k)$. Explicitely
\begin{eqnarray}
\Lambda(\gamma,R) = \Lambda_1(\gamma)/R^{k} + O(1/R^{k+1})\, ,
\end{eqnarray}
where $\Lambda_1(\gamma)\uparrow +\infty$ as $\gamma\uparrow\gamma_{\rm s}(k)$.
 
In the following we shall avoid dealing with the above phenomenon by
focusing directly on the unfrustrated ensemble: this will enable us to use 
efficient linear algebra techniques for numerical computations. 
There are several justifications for doing this:
\begin{enumerate}
\item The two ensembles become equivalent in the Kac limit, 
which is our main concern here.
\item We are interested in the long distance properties of the model, rather 
than in the effect of small substructures. We think that the
two decouple for large $R$.
\item Even if the frustrated ensemble is with high probability unsatisfiable,
one can always consider `almost satisfying' configurations.
Equivalently, one can study the Boltzmann measure for the energy
(\ref{eq:Energy}) at a small non-vanishing temperature $T$.
We expect the effect of small frustrated substructures on the 
thermodynamics to be small, and indeed vanishing as $R^{-k}$
for large $R$.
\end{enumerate}
In this perspective, we shall introduce 
an \emph{improved ensemble} which reduces the effect
of small substructured, while keeping 
the large $R$ behavior unchanged.
 This is particularly convenient in numerical simulations.

\begin{figure}[t]
\includegraphics[width=0.575\linewidth]{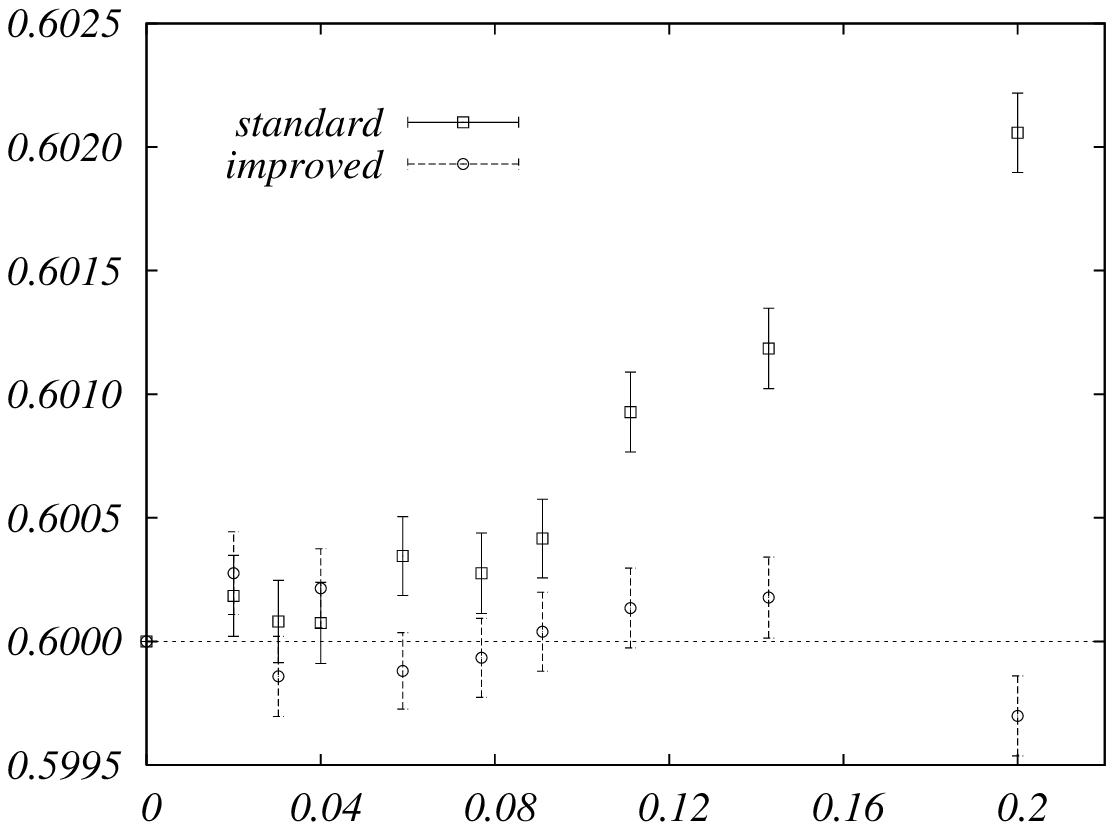}
\put(-130,-7){$1/R$}
\put(-290,100){$\phi_R(\gamma)$}
\caption{Ground state entropy density in the thermodynamic limit
$\phi_R(\gamma) = \lim_{L\to\infty} \phi_{L,R} ( \gamma )$,
cf. Eq.~(\ref{eq:EntropyDensityDef}), for the standard and improved 
ensembles. Here $k=3$ and $\gamma = 0.4$. The horizontal line marks the 
$R\to\infty$ limit $\phi(\gamma) = 0.6$.}
\label{fig:standard_vs_improved}
\end{figure}
Ideally, one would like 
to consider a uniform ensemble conditioned on some class of substructures
being absent. 
In practice it can be excedingly difficult to sample matrices $\H$
from such a conditional ensemble. We shall define the improved ensemble 
by the following sequential procedure. First generate the 
random set $F\subseteq [L]$ by letting $i\in F$ independently for each 
$i=1,\dots, L$ with probability $\gamma$. The set $F$ will index lines of
$\H$ as above. Then we choose a uniformly random ordering
$(i(1),\dots, i(M))$ of the elements of $F$,
and generate the corresponding lines of $\H$ 
following such an order. For each $t=1,\dots, M$ 
we try to generate the line of $\H$ indexed by $i(t)$ by drawing its $k$
non-zero elements uniformly at random in $\{i(t)-R,\dots,i(t)+R\}$.
If the newly generated line has $k-1$ or $k$ non-vanishing 
entries in common with one of the previously generated lines
$\{i(1),\dots,i(t-1)\}$, we reject it and re-sample it.
We repeat the trial-rejection step for at most $100$ times.
If no valid line is generate within this round, the whole system generated
so-far is rejected and the procedure is re-initiated from scratch.

We shall refer to the first ensemble introduced above as to the
\emph{standard}, whenever it will be necessary to distinguish it from the 
\emph{improved} ensemble. In Fig.~\ref{fig:standard_vs_improved}
we compare the $R\to\infty$ behavior of the ground state entropy
for the two ensembles. Although the limits clearly coincide, the improved 
ensemble is close to it even for $R=5$. 
%
%
\section{Ground state entropy}
\label{sec:GroundState}

The simplest thermodynamic quantity that is relevant for
the study of the unfrustrated KacXOR problem is the ground state entropy, i.e.
the logarithm of the number of solutions of the linear system.
Let us denote by $Z(\H)$ the number of solutions of
the linear system (\ref{eq:XORformula}) for a random binary 
matrix $\H$. 
Then the average entropy density is defined as
\begin{equation}
 \phi_{L,R} ( \gamma ) = \frac{1}{L} \mathbb{E} \log_2 Z(\H)\,,
\label{eq:EntropyDensityDef}
\end{equation}
In order to compare analytical predictions and numerical data
it will be convenient to define the `subtracted' entropy density
$\ph_{L,R}(\gamma) \equiv \phi_{L,R}(\gamma)-\phi_{\rm naive}(\gamma)$,
where $\phi_{\rm naive}(\gamma) = 1-\gamma$.
Notice that $\phi_{\rm naive}(\gamma)$ is the naive prediction
that would be obtained by assuming the lines of $\H$ to be linearly 
independent.

Given a matrix $\H$, the corresponding number of solutions takes the 
form of a partition function
\begin{eqnarray}
Z(\H) = \sum_{\ux}\prod_{a=1}^L\psi_a(x_{a -R},\dots,x_{a+R})\, ,
\label{eq:PartFun}
\end{eqnarray}
where (denoting by $\oplus$ the sum modulo $2$)
\begin{equation}
\psi_a(x_{a-R},\ldots,x_{a+R}) = \left| \begin{array}{ll}
\ind (x_{i_1(a)} \oplus \ldots \oplus x_{i_k(a)}=0) & \;\;\mbox{if $a\in F$,} \\
1 & \;\;\mbox{if $a\not\in F$.}
\end{array} \right.
\end{equation}
Due to the finite interaction range $R$, $Z = Z(\H)$ can be 
computed through a transfer matrix algorithm which recursively computes
left and right partition functions, respectively $Z_{\to i}$ and
$Z_{i\ot}$. These are indexed by 
$\vz=(z_1,\dots,z_{2R})\in\{\0t,\1t\}^{2R}$, and defined as
\begin{eqnarray}
Z_{\to i}(\vz) &\equiv& \sum_{\substack{x_1\dots x_i \\
\vx_{i-2R+1}^{i}=\vz}} \prod_{a=1}^{i-R}\psi_a(x_{a -R},\dots,a+R)\, ,
\label{eq:ConstrPartFun}\\
Z_{i\ot}(\vz) &\equiv& \sum_{\substack{x_i\dots x_i\\
\vx_{i}^{i+2R-1}=\vz}} \prod_{a=i+R}^{L}\psi_a(x_{a -R},\dots,a+R)\, ,
\end{eqnarray}
where we used the notation $\vx_{j+1}^{j+2R} = (x_{j+1},\dots,x_{j+2R})$.
A recursion naturally follows
\begin{eqnarray}
Z_{\to (i+1)}(z_1,\dots,z_{2R})  =\sum_{z_0\in\{\0t,\1t\}} 
\psi_{i-R+1}(z_0,z_1,\dots,z_{2R})\,
Z_{\to i}(z_0,\dots,z_{2R-1})\, ,\label{eq:RecPartFun}
\end{eqnarray}
together for the analogous recursion for $Z_{i\ot}$. 
It is clear that the total number of solutions can be computed from the 
constrained partition functions.

The naive transfer matrix algorithm defined by the recursion 
(\ref{eq:RecPartFun}) has complexity that of order 
$\Theta(L2^{2R})$. This severely limits the interaction ranges that 
can be treated with this method: in practice we could deal at most
with $R= 10\div 11$, which is far too small to address issues concerning the
$R\to\infty$ limit. In order to overcome this problem, we developed
a transfer matrix algorithm that, while computing exactly
the  constrained partition functions, exploits the linear 
structure of the problem in such a way to reduce the complexity to
$\Theta(LR^3)$. Thanks to this approach, we were able to treat systems with 
$R=100$ or larger. For details on the algorithm we refer to
Appendix \ref{app:Algo}.

We are interested in the double limit $R,L\to\infty$. We shall consider 
two procedures to define the limit. The first one corresponds to the classical
Kac limit, and consists in taking the thermodynamic limit upfront
to define
\begin{equation}
 \phi_R ( \gamma ) \equiv  \lim_{L \to \infty} \phi_{L,R} ( \gamma ) \, .
\end{equation}
Next, we let $R\to \infty$. In Appendix \ref{app:Analytical} we will
show that $\phi_{R}(\gamma)$ can be expanded for large $R$ as
follows
\begin{equation}
\phi_R ( \gamma ) = \phi^{(0)} (\gamma) + \frac{1}{2 R +1}\, \phi^{(1)} (\gamma) + O \left( \frac{1}{R^2} \right)\, . \label{eq:AsymptoticMF}
\end{equation}
The leading term gives the mean-field limit and coincides with the
ground state entropy density within the rXOR ensemble
\cite{XOR_1,XOR_2}.
It can be expressed in the form
$\phi^{(0)} (\gamma) = \max_{\varphi\in[0,1]}
\phi^{(0)} (\gamma;\varphi)$, where
\begin{equation}
\phi^{(0)} (\gamma;\varphi) = - \gamma ( 1 - \varphi^k ) + k \gamma \varphi^{k-1}(1 - \varphi) + e^{- k \gamma \varphi^{k-1} } \, .\label{eq:MFFreeEnergy}
\end{equation}
It is easy to show that the $\max$ is achieved for a value of the order
parameter $\varphi$ that satisfies the equation $\varphi 
= 1 - \exp\{-k\varphi^{k-1}\}$. For $\gamma<\gamma_{\rm s}(k)$,
the maximum is at $\varphi = 0$, yielding $\phi^{(0)}(\gamma) = 1-\gamma$.
In other words, the the rank of $\H$ is smaller than the maximum
possible value by a fraction of order $1/R$.
For $\gamma\ge\gamma_{\rm s}(k)$, the maximum is at $\varphi=\varphi_*>0$
strictly, and  $\phi^{(0)}(\gamma) > 1-\gamma$:
the rank of $\H$ remains strictly smaller than its maximum possible 
value even as $R\to\infty$.
For instance we have $\gamma_{\rm s}(k) \approx 0.917935$
for $k=3$.

The first-order contribution $\phi^{(1)}(\gamma)$ is related to 
fluctuation around the saddle point in an appropriate path integral 
representation of Eq.~(\ref{eq:PartFun}). Its expression 
is given in Appendix \ref{app:Analytical}.

\begin{figure}[t]
\includegraphics[width=0.7\linewidth]{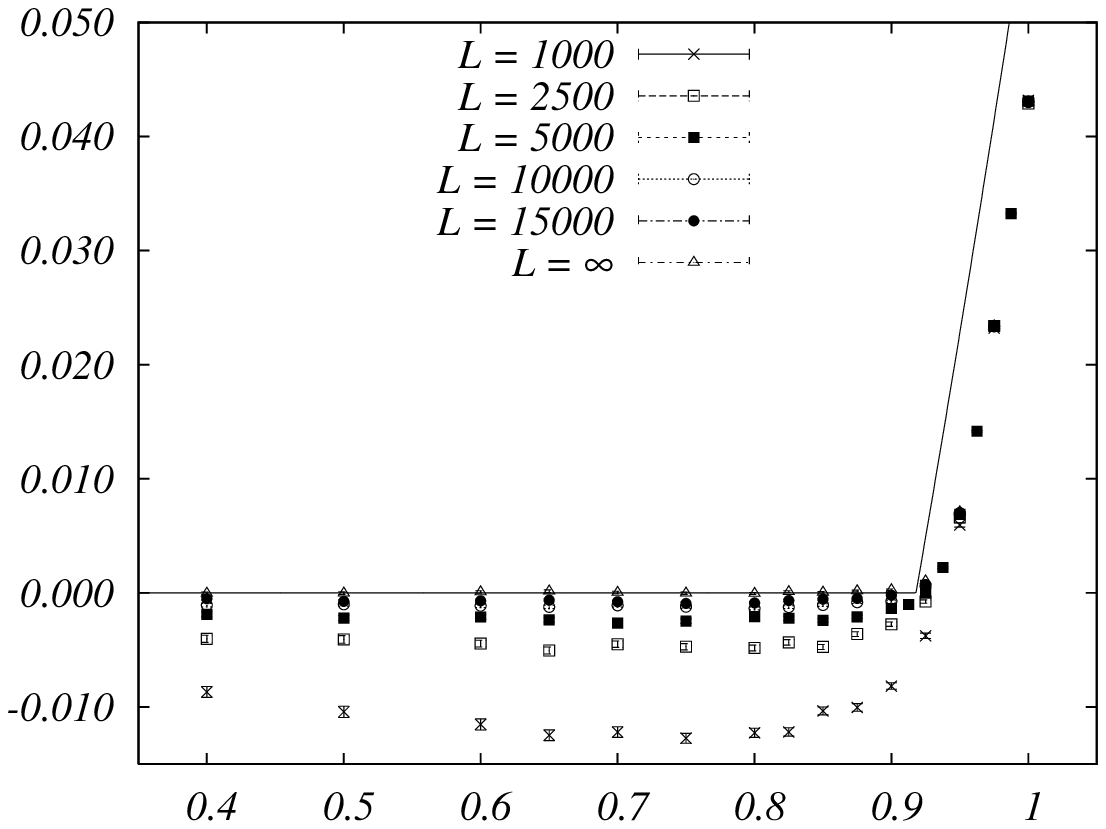}
\put(-150,-7){$\gamma$}
\put(-340,120){$\ph_{L,R}(\gamma)$}
\caption{Subtracted entropy density
$\ph_{L,R}(\gamma)=\phi_{L,R}(\gamma)-1+\gamma$ 
for various values of $L$ and $R=25$ constant.
We also plot the result of an $L\to\infty$ extrapolation, 
and the analytical mean-field prediction $\phi^{(0)}(\gamma)-1+\gamma$ 
(continuous line).}
\label{L_ext_R25_gs}
\end{figure}
In Fig.~\ref{L_ext_R25_gs} we plot the numerical estimates for the
subtracted entropy density $\ph_{L,R}(\gamma)$, as obtained with our 
transfer matrix algorithm for $R=25$ and several system sizes.
Data points are the result of averaging over $1000$ realizations
of $\H$ with $k=3$. The same statistics and value of $k$
will be used in the other numerical experiments below: we shall omit
mentioning it again.
Further, unless otherwise stated, we will keep using the improved ensemble.
We also show the result of an $1/L$ extrapolation to $L=\infty$.
The control of the thermodynamic limit is quite good (although 
corrections at moderate values of $L$ are large). It is clear that the 
$L=\infty$ extrapolation is not compatible with the mean field prediction,
and that $1/R$ corrections must be taken into account.

\begin{figure}[t]
\includegraphics[width=0.7\linewidth]{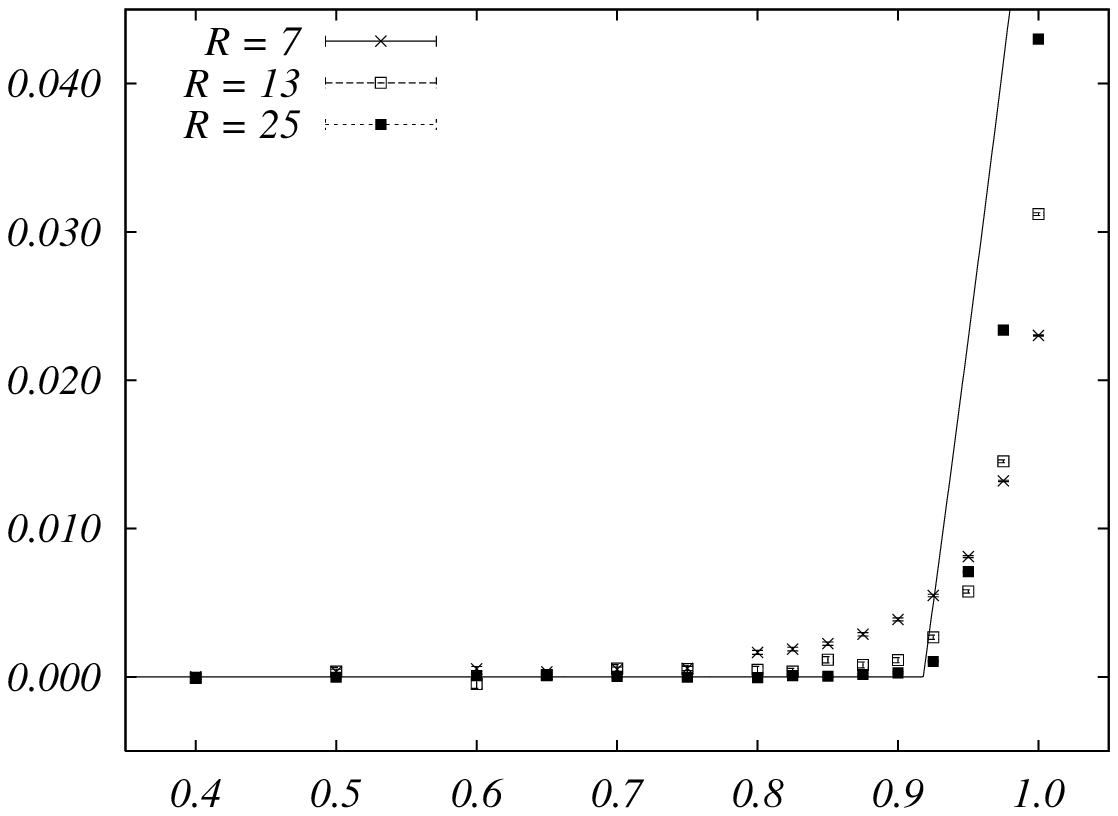}
\put(-150,-7){$\gamma$}
\put(-340,120){$\ph_{R}(\gamma)$}
\caption{Subtracted entropy density in the thermodynamic limit:
$\ph_{R}(\gamma)=\phi_{R}(\gamma)-1+\gamma$ for various values of $R$,
together with the mean field prediction  $\phi^{(0)}(\gamma)-1+\gamma$ 
(continuous line).}
\label{evol_R_gs_L5000}
\end{figure}
\begin{figure}[h!]
\includegraphics[width=0.49\linewidth]{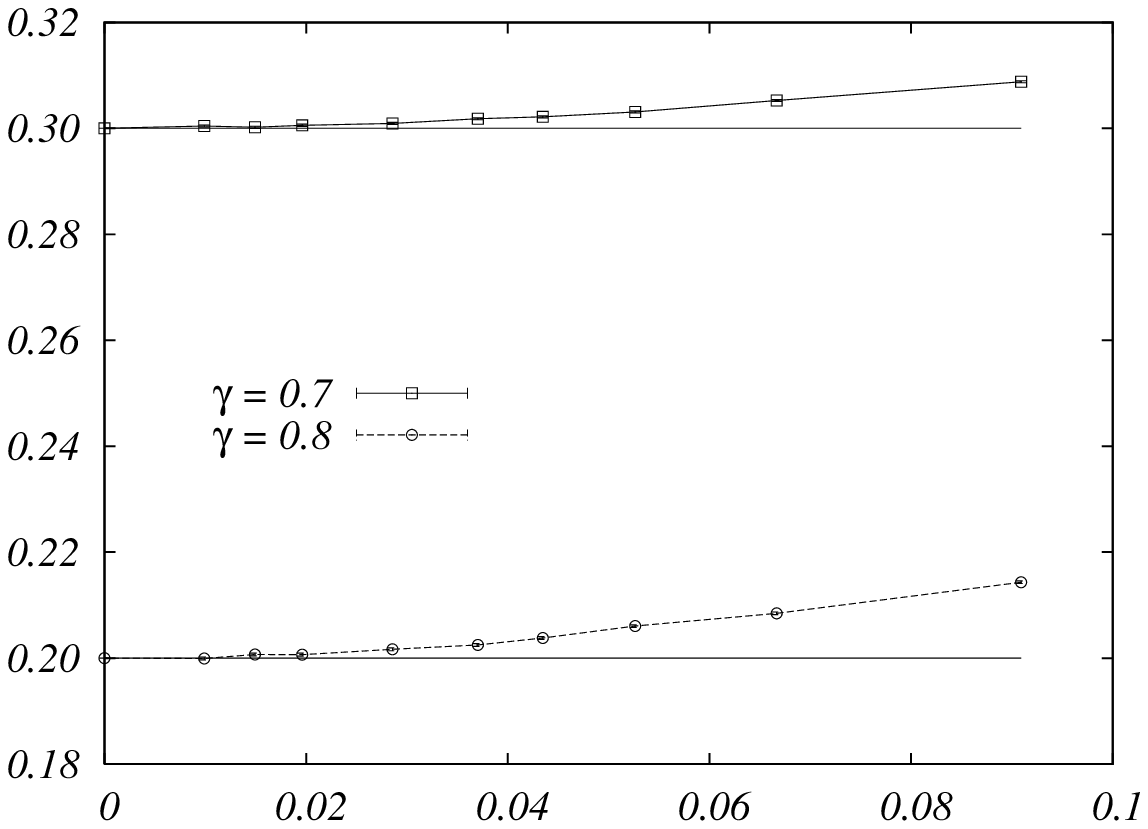} 
\includegraphics[width=0.49\linewidth]{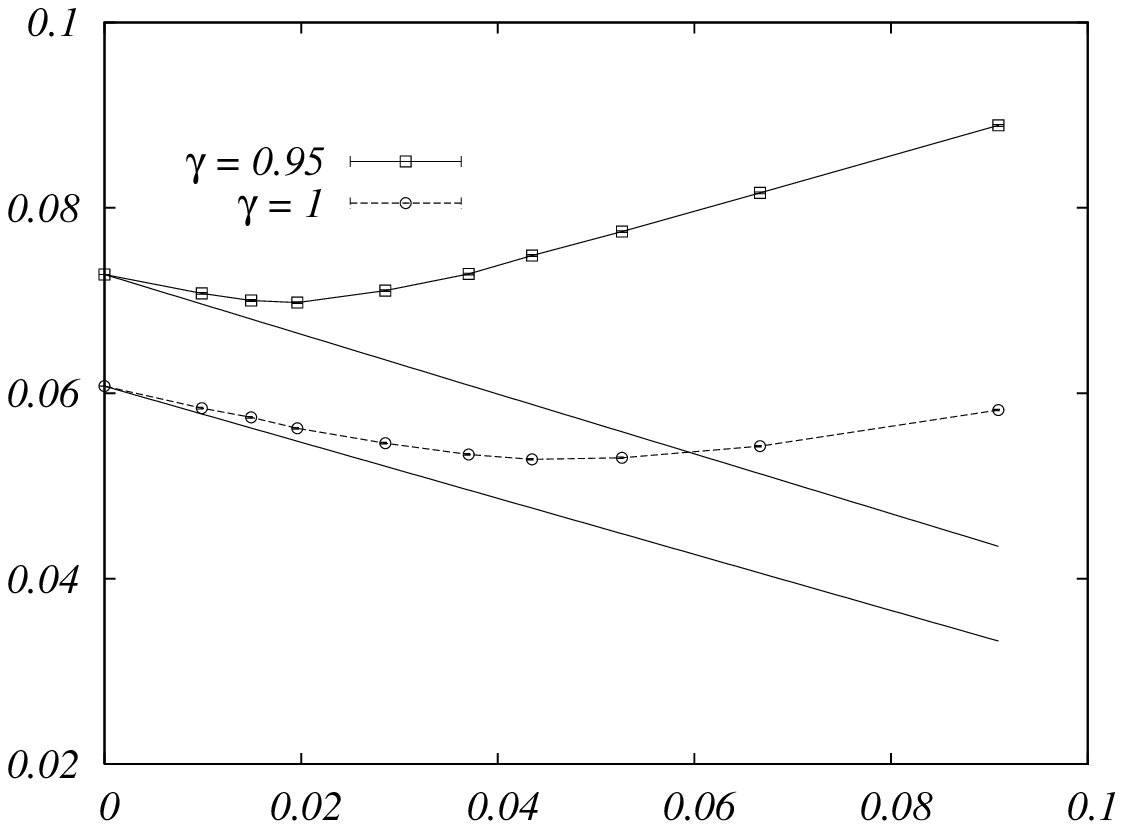}
\put(-370,-8){$1/(2R+1)$} 
\put(-120,-8){$1/(2R+1)$}
\put(-480,90){$\phi_R(\gamma)$} 
\put(-235,90){$\phi_R(\gamma)$} 
\caption{Ground state entropy density $\phi_R(\gamma)$,
extrapolated to the thermodynamic limit, versus the inverse
interaction range $1/(2R+1)$ for various 
values of $\gamma$. Straight lines correspond to the 
analytic prediction, cf. Eq.~(\ref{eq:AsymptoticMF}).}
\label{vs_r_L5000_gs}
\end{figure}
Figure \ref{evol_R_gs_L5000} shows the result of such an $L\to\infty$
extrapolation for several values of $R$. The data seem to approach the
mean field prediction $\phi^{(0)}(\gamma)-1+\gamma$ as $R\to\infty$,
although the approach is rather slow. 

In order to better study the large-$R$ limit, for $4$ values of
$\gamma$ we computed the ground state entropy for a wide range of $R$. 
The result is compared in Figure \ref{vs_r_L5000_gs} with the asymptotic 
expression (\ref{eq:AsymptoticMF}). In this case we used 
the standard ensemble which presents larger $1/R$ corrections
(computing the first order correction $\phi^{(1)}$ within the 
improved ensemble is technically much more difficult).
It turns out from the analysis in Appendix \ref{app:Analytical}
that $\phi^{(1)}(\gamma) = 0$ for $\gamma<\gamma_{\rm s}(k)$
while $\phi^{(1)}(\gamma) \neq 0$ for $\gamma\ge\gamma_{\rm s}(k)$.
Our data confirm this behavior. Further, although $O(R^{-2})$ contributions 
are rather large, the leading $1/R$ correction to mean field
does indeed match the analytical prediction.

The second limit we shall consider is $L, R\to\infty$
with $\ell \equiv L/R$ kept fixed. We thus define
\begin{eqnarray}
\phi^*_\ell(\gamma) \equiv\lim_{R\to\infty} \phi_{R\ell,R}(\gamma)\, .
\label{eq:FSMeanField}
\end{eqnarray}
This is the mean-field limit for a system of finite size. 
The limit can be computed exactly by maximizing an appropriate
action functional over a position-dependent order parameter.
More precisely we have $\phi^*_\ell(\gamma) = \max_{\varphi} A[\varphi]$,
where $\varphi:[0,\ell]\to \reals$ is the order parameter,
and 
\begin{eqnarray}
A[\varphi] = \frac{1}{\ell}\int_0^{\ell}
\left\{\gamma-k\gamma\varphi(z)^{k-1}+(k-1)\gamma\varphi(z)^{k}
-\exp\left[-\frac{k\gamma}{2}\int_{-1}^{1} \varphi(z+u)^{k-1}\de u\right]\right\}
\, \de z\, .\label{eq:Action}
\end{eqnarray}
By differentiating with respect to $\varphi$, we obtain the mean-field equation
\begin{eqnarray}
\varphi(z) = 1-\frac{1}{2}\int_{-1}^{+1}\exp\left\{
-\frac{k\gamma}{2}\int_{-1}^{+1}\varphi(z+u+v)^{k-1}\de v\right\}\de u\, .
\label{eq:FiniteRangeMFEq}
\end{eqnarray}
We refer to Appendix \ref{app:Analytical} for a derivation 
of these formulae and limit ourselves to discuss their consequences here.

In the case of a homogeneous order parameter $\varphi(z) =\varphi$
independent of $z$, Eq.~(\ref{eq:FiniteRangeMFEq}) is satisfied if 
$\varphi$ if a solution of the standard mean field equation,
$\varphi = 1-\exp\{-k\gamma\varphi^{k-1}\}$. The action 
(\ref{eq:Action}) then reduces to the mean field free 
energy Eq.~(\ref{eq:MFFreeEnergy}). 

In the general case the order parameter $\varphi(z)$ has a simple 
interpretation. Consider the linear system $\H\ux = \u0t$
$\mod 2$, and let $i\in\{1, \dots,L\}$. Then, one of the following
must happen: either $x_i=\0t$ in all of the solutions;
or $x_i=\0t$ in half of the solutions and $x_i=\1t$ in the other
half.  We shall call $x_i$ (or, sometimes, $i$) a frozen variable in the 
first case, and a free variable in the second one. 
Given $z\in[0,\ell]$, the number of frozen variables 
with $i\in [Rz,R(z+\de z)]$ 
in a typical random linear system from our ensemble,
is about $R\varphi(z)\, \de z$. Equivalently, the probability for $x_i$,
$i = \lfloor Rz\rfloor$, to be frozen converges to $\varphi(z)$.

We shall come back to this interpretation in the next Section,
while using it here to derive the appropriate boundary conditions for 
Eq.~(\ref{eq:FiniteRangeMFEq}). If the linear system is defined with 
periodic boundary conditions, then we have to use periodic boundary 
conditions in Eq.~(\ref{eq:FiniteRangeMFEq}) as well, namely
$\varphi(z+\ell) = \varphi(z)$. If on the other hand we adopt
fixed boundary conditions with respect to the reference solution
$\ux^{(0)} = \u0t$, then we have to impose  $\varphi(z)=1$
for $z\le 0$ and $z\ge \ell$ in  Eq.~(\ref{eq:FiniteRangeMFEq}).
As a consequence, the homogeneous solution is no longer relevant in this case.

Once the boundary conditions have been estabilished, 
Eq.~(\ref{eq:FiniteRangeMFEq}) can be solved numerically by iteration 
(after discretizing $z$ on a sufficiently fine mesh). In the regime in
which multiple fixed points exist, the relevant one is obtained by 
maximizing the action.
\begin{figure}
\includegraphics[width=0.7\linewidth]{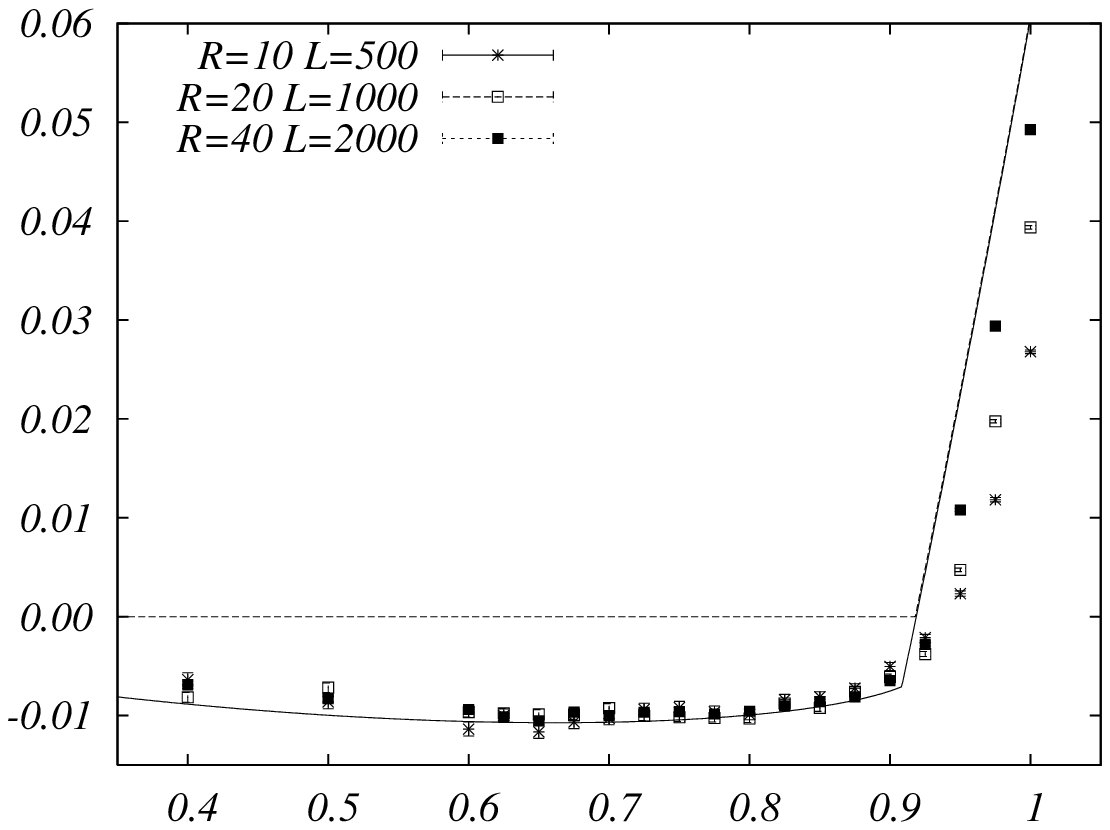}
\put(-150,-5){$\gamma$}
\put(-340,120){$\ph_{L,R}(\gamma)$}
\caption{Subtracted entropy density 
$\ph_{L,R}(\gamma) = \phi_{L,R}(\gamma)-1+\gamma$ as a function of
$\gamma$ for several at $\ell = L/R = 50$ fixed.
The continuous line corresponds to the analytical prediction
$\phi^*_\ell(\gamma)-1+\gamma$ in the $R\to\infty$ limit.}
\label{gs_fixed_l}
\end{figure}

The result of such a computation is compared in Fig.~\ref{gs_fixed_l}
with the outcome of numerical simulations. The agreement is good
already at moderate interaction ranges. The main effect of a finite 
size is a decrease in the number of solutions due to the fact that
variables close to the boundary are more highly constrained
(and thus more likely to be frozen). This effect is accurately reproduced by 
the analytical calculation.
%
%
\section{Point-to-set correlation function}
\label{sec:PointToSet}

As we have seen in the previous Section, the thermodynamic behavior
of the KacXOR ensemble at finite $R$ carries several traces of the
mean field limit.  Here we want to investigate some structural features
of the uniform measure over solutions of the linear
system:
\begin{eqnarray}
\mu(\ux) = \frac{1}{Z}\, \ind(\H\,\ux = \u0t) \equiv
\frac{1}{Z}\, \prod_{a}\psi_a(x_{a-R},\dots,x_{a+R})\, .
\end{eqnarray}
In particular, we want to understand whether the
mean field ergodicity breaking transition shows up in 
the long range correlations of this measure, as predicted within the 
mosaic state scenario.

It is expected that the long range order emerging at a glass transition 
cannot be probed through ordinary point-to-point correlations functions,
and that point-to-set correlation functions have to be used instead
\cite{MonSem}. 
These can be defined through the following ``experiment'' 
\cite{BB2} 
(we refer here to the one-dimensional case we are studying).
Consider a large sample $L\gg R$, let $i$ be a node in its bulk:
$i\gg R$, $L-i\gg R$, and $\ux^{*}$ a `reference' configuration sampled 
from the measure $\mu(\, \cdot\, )$. Then fix  some $1\le\tL\ll L$, and 
consider a second 
configuration that is forced to coincide with 
$\ux^{*}$ on sites $j$ with $|j-i|> \tL$, and free otherwise,
and compute the probability that $x_i\neq x_i^*$. 
The expectation of this probability with respect to $\ux^*$ and
the sample realization yields the desired point-to-set correlation.
In formulae,
if we let $\ball(i,\tL)=\{j:\, |i-j|\le \tL\}$ be the box
of size $2\tL+1$ around $i$,
and $\cball(i,\tL)$ its complement, we define 
\begin{eqnarray}
\widetilde{G}(\tL,R,\gamma) \equiv \E\{1-2\mu_{i|\cball(i,\tL)}(x_i\neq x_i^*|\ux^*_{\cball(i,\tL)})\}\, . \label{eq:PointToSet}
\end{eqnarray}
Here the thermodynamic limit $L\to\infty$ is assumed to be taken at the outset,
$\E$ denotes expectation both with respect to the matrix $\H$ and the 
reference configuration $\ux^*$, and the redefinition 
$1-2(\cdots)$ is for future convenience. 

\begin{figure}[t]
\includegraphics[width=0.6\linewidth]{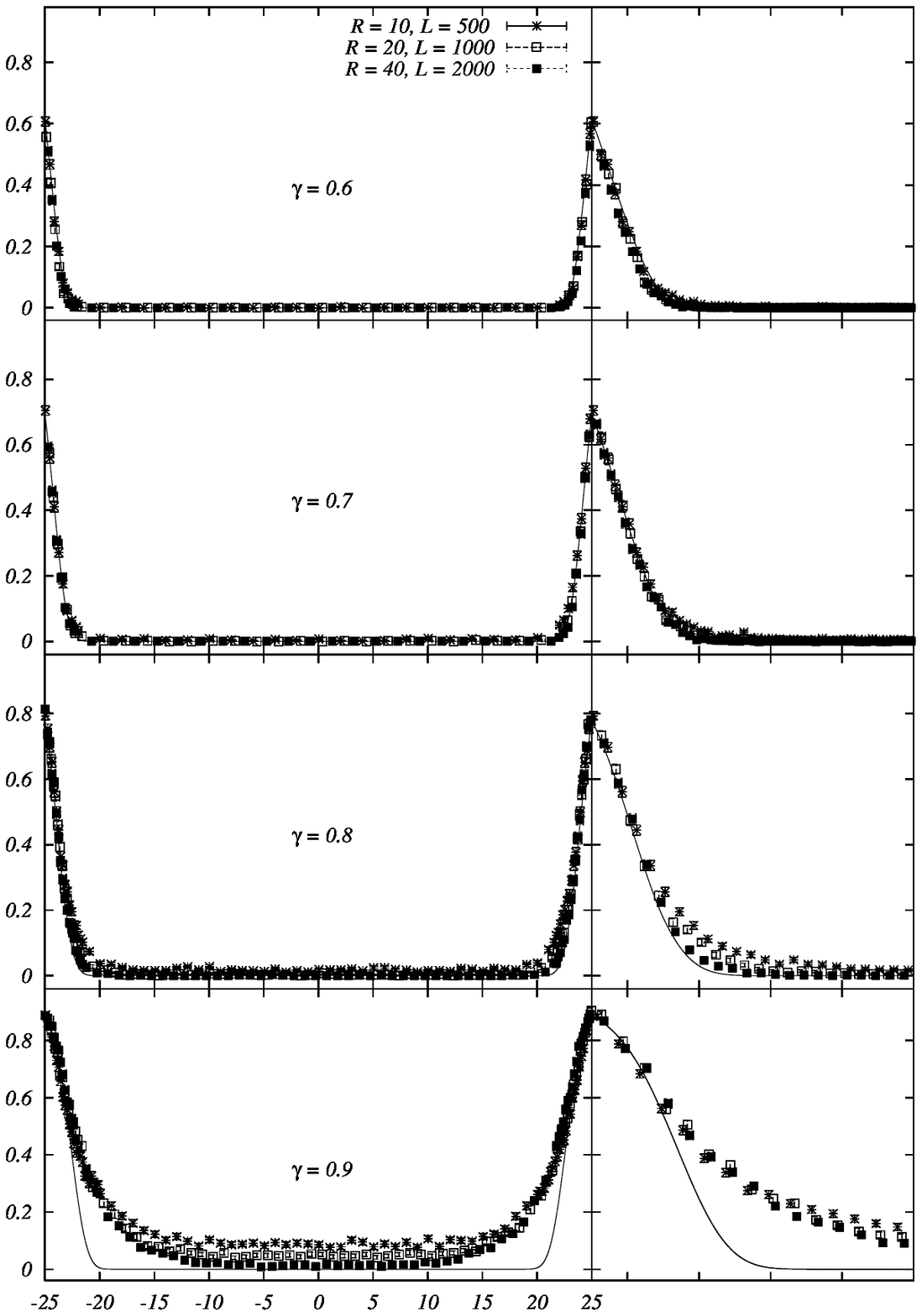}
\put(-150,-5){$z$}
\put(-290,200){$\widetilde{G}$}
\caption{Correlation $\widetilde{G}(n;\tL,R,\gamma)$ between the boundary
of a box of size $2\tL+1 = 2\ell R+1$ and a point in its interior
(at distance $n=zR$ from the center). In the right frames: blow-up of
the region near the boundary. The continuous line (partially hidden by 
data points) corresponds to the analytic prediction obtained by solving 
Eq.~(\ref{eq:FiniteRangeMFEq}).}
\label{fig:profiles}
\end{figure}
\begin{figure}[t]
\includegraphics[width=0.6\linewidth]{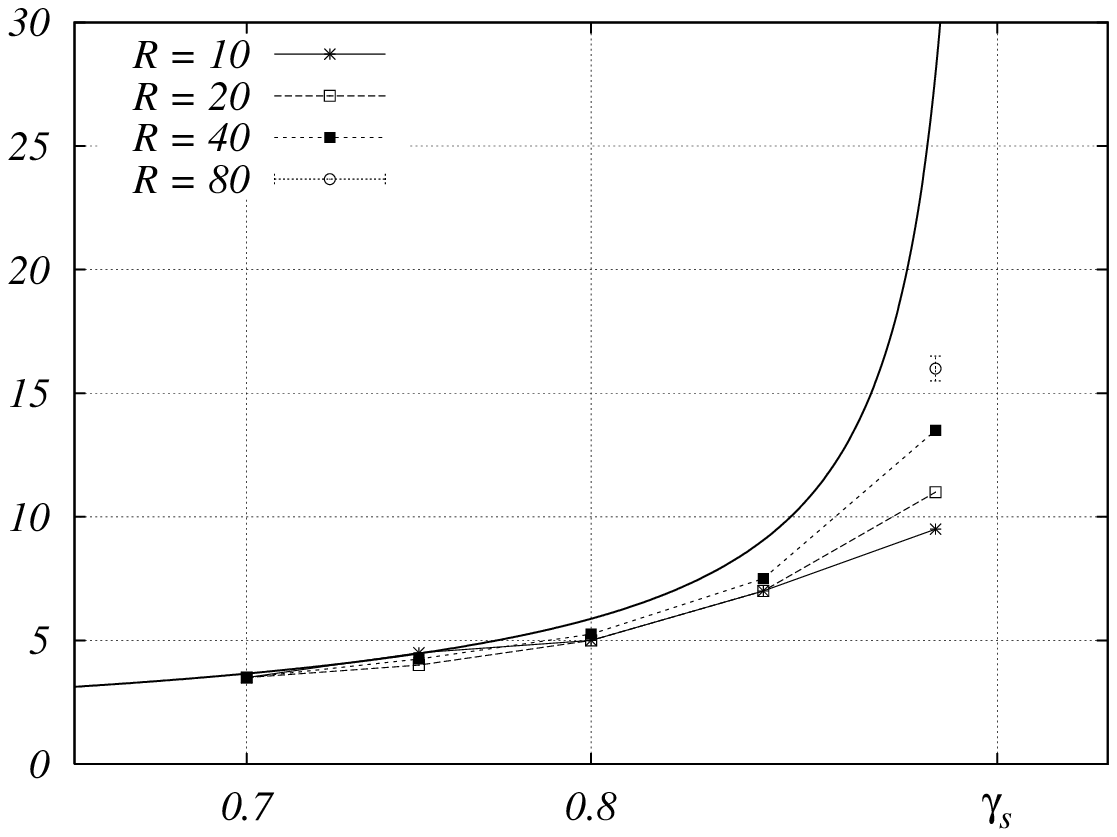}
\put(-140,-5){$\gamma$}
\put(-280,100){$\ell_{\rm s}$}
\caption{Point-to-set correlation length in units of the interaction range $R$.
The continuous line corresponds to the analytic prediction for $R\to\infty$
and diverges at the glass transition 
$\gamma_{\rm s}(k=3)\approx 0.917935$.}
\label{fig:xi_vs_gamma}
\end{figure}
The linear structure of our problem implies two simplifications.
First, the conditional probability appearing in Eq.~(\ref{eq:PointToSet})
is indeed independent of $\ux^*$ (that can be `gauged away').
Therefore we can fix $\ux^*=\u0t$ and eliminate the expectation 
over the reference configuration $\ux^*$.
The resulting conditional measure is just the distribution of
a system with fixed boundary conditions $\u0t$ as discussed 
in the previous Section. This implies a second simplification (already 
noticed above).
The conditional probability
$\mu_{i|\cball(i\tL)}(x_i\neq \0t|\ux^*_{\cball(i\tL)}=\u0t)$
can take value $1/2$ (if $x_i$ is `free') or $0$ (if it is `frozen').
We thus get 
\begin{eqnarray*}
\widetilde{G}(\tL,R,\gamma) =\prob_{\tL}\{\, x_{\tL+1} \mbox{ is free}\, \}\, .
\end{eqnarray*}
Here $\prob_{\tL}$ denotes probability with respect to a matrix
$\H$ with $2\tL+1$ columns and fixed $\0t$ boundary conditions.
In fact it is interesting to generalize the above definition and
consider the correlation between any point inside the  box
of size $2\tL$ and its boundary
\begin{eqnarray*}
\widetilde{G}(n;\tL,R,\gamma) \equiv \prob_{\tL}\{\, x_{\tL+1+n} \mbox{ is free}\, \}\, .
\end{eqnarray*}
The original definition is recovered for $n=0$.

We expect $\widetilde{G}(n;\tL,R,\gamma)$ to be close to $1$
when $n$ approaches the boundaries of the box 
(i.e. $n\approx \tL$ or $n\approx -\tL$) and to decrease in the interior.
If the box is large enough, it will approach its thermodynamic value,
independent of the boundary condition, near the center (for $n\approx 0$). 
In Figure \ref{fig:profiles} we show the outcomes of a numerical calculation
of $\widetilde{G}(n;\tL,R,\gamma)$ for several values of its parameters.

We are particularly interested in the mean field limit. This is obtained 
by defining
\begin{eqnarray}
G(z;\ell,\gamma) \equiv \lim_{R\to\infty} \widetilde{G}(n=Rz;\tL=R\ell,R,
\gamma) \, ,
\end{eqnarray}
that is by measuring lengths in terms of the interaction range and 
letting $R\to\infty$. In agrement  with the interpretation 
of the previous section, we expect $G(z;\ell,\gamma) =\varphi(z)$,
where $\varphi(z)$ solves Eq.~(\ref{eq:FiniteRangeMFEq}) 
with boundary condition
$\varphi(z) = 1$ for $z\le -\ell$, and for $z\ge \ell$.
The comparison with numerical data in Fig.~\ref{fig:profiles}
is satisfactory although the convergence to the $R\to\infty$
limit gets slower and slower as $\gamma_{\rm s}(3)\approx 0.917935$ 
is approached.

The point-to-set correlation function can be used to
define a correlation length, namely the smallest box size such that
the correlation is below a pre-estabilished constant $\varepsilon$.
Here we will choose\footnote{Any strictly positive constant
below the Edwards-Anderson parameter (in this case given by the
largest solution of $\varphi = 1-\exp\{-k\gamma\varphi^{k-1}\}$)
should provide an equivalent definition.} $\varepsilon = 1/2$.
In formulae 
\begin{eqnarray}
\ell_{\rm s}(\gamma,R) = 
\min\{\;\ell\, : \; \widetilde{G}(\tL=R\ell,R,
\gamma)\le 1/2\;\}\, .
\end{eqnarray}
An analytical prediction in the $R\to\infty$ limit 
can be obtained by solving  Eq.~(\ref{eq:FiniteRangeMFEq}) with
boundary conditions $\varphi(z) = 1$ for $z\not\in[-\ell,\ell]$.
The resulting length can be shown to diverge at $\gamma_{\rm s}$ 
as $\ell_{\rm s}(\gamma,R=\infty) \sim(\gamma_{\rm s}-\gamma)^{-1}$,
in agreement with the mosaic picture (indeed 
$\Sigma(\gamma)\sim (\gamma_{\rm s}-\gamma)$ close to the transition).

In Fig.~\ref{fig:xi_vs_gamma} we compare this prediction with 
the estimates from numerical simulations at finite $R$. These
two are clearly consistent, although the convergence is rather 
slow in the critical regime.
%
%
\section{Discussion}
\label{sec:Discussion}

We defined a simple ensemble of constraint satisfaction problems 
(more precisely, an ensemble of linear problems over integers modulo 
$2$), with 
one-dimensional Kac structure. The model is exactly soluble
for infinite interaction range
$R\to\infty$ and exhibits in this limit a glassy phase with an exponential 
number of pure states and a SAT-UNSAT transition.

Mean field theory (as interpreted within the mosaic picture)
seems to describe the behavior of the system at moderately large $R$.
Indeed we were able to get quantitative predictions by taking
into account the principal modifications of naive mean-field theory,
namely a position-dependent order parameter, and $1/R$
corrections.
In particular we checked for the first time the divergence 
of the mosaic length scale in a concrete model, by comparing the
the result of a controlled approximation (large $R$ limit)
with exact numerical calculations.

We think the KacXOR model can be a useful playground for many
ideas developed in the physics of glasses. Among several interesting
research directions, one might consider: $(i)$ Studying the frustrated 
ensemble (corresponding to an inhomogeneous linear system);
$(ii)$ Introducing a non-vanishing temperature and studying the
corresponding Boltzmann distribution;
$(iii)$  Studying the behavior of  Glauber dynamics, and in particular
the relation between relaxation time and mosaic length scale.

On a different theme, ensembles of random constraint satisfaction
problems have been recurrently used to test heuristic algorithms
\cite{XOR_CS}.
Such tests have limited scope because in practical applications instances
are often structured. It might be insightful therefore to
consider ensembles with some tunable `structure parameter',
such as the interaction range $R$ in the present model.
%
%
\appendix

\section{Counting small substructures}
\label{app:Janson}

Consider the random linear system $\H\ux =\ub$ defined in Section
\ref{sec:Definition}. If two lines $i,j\in F$ in $\H$
are equal, while the corresponding entries in $\ub$ (namely $b_i$ and $b_j$)
are different, then the system has no solution. We call such a pair
$(i,j)$ a `bad pair,' and will write $B_{ij} = 1$ if $(i,j)$ is bad,
and $B_{ij}=0$ otherwise. Therefore
\begin{eqnarray}
\prob\left\{\H\ux=\ub \mbox{ is SAT}\right\}\le 
\prob\{\cap_{(i,j)} [B_{ij}=0]\}\, ,
\end{eqnarray}
where the intersection ranges over $i,j$ such that $i<j\le i+2R+1-k$.
Let $B = \sum_{(ij)}B_{ij}$ be the number of bad pairs.
In order to bound the right hand side above, we use 
Janson's inequality \cite{AlonSpencer}, which implies
\begin{eqnarray}
\prob\left\{\H\ux=\ub \mbox{ is SAT}\right\}\le 
\exp\left\{-\E[B] + \Delta/2(1-\epsilon)\right\}\, .
\end{eqnarray}
Here 
\begin{eqnarray}
\epsilon = \sup_{(ij)}\, \E[B_{ij}]\, ,\;\;\;\;\;\;\;
\Delta = \sum_{(ij)\sim (lm)}\E[B_{ij}B_{lm}]\, .
\end{eqnarray}
where the sum over $(ij)\sim (lm)$ runs over all the couples
of distinct pairs $(ij)$ and $(lm)$ such that $B_{ij}$ and $B_{lm}$
are not independent.

It is easy to realize that both $\E[B]$ and $\Delta$ 
are of order $\Theta(L)$ since they are sums of $\Theta(L)$ 
positive terms. 
Since we are only interested in the coefficient of
the order $L$ term, we shall always consider pairs $(ij)$ in the bulk.
Then we have
\begin{eqnarray}
\E[B_{ij}] = \frac{\gamma^2}{2{2R+1 \choose k}^2} {2R+1-|i-j| \choose k}\, .
\end{eqnarray}
The factor $\gamma^2$ has to be included for having $i,j\in F$
(the two equations must present), $ {2R+1-|i-j| \choose k}/{2R+1 \choose k}^2$
is the probability that the two lines in $\H$ coincide, and $1/2$
is the probability that $b_i\neq b_j$

Since the above expression is maximized for $|i-j|=1$,
we have $\epsilon = K_0\gamma^2$, with $K_0$ as in Eq.~(\ref{eq:K0K1}).
Further, by summing over $i,j$ we obtain $\E[B] = K_1\gamma^2 L+O(1)$
for $L\to\infty$.

As for the term $\Delta$, the only non-vanishing contribution
comes form the case in which there are three distinct 
indices among $\{i,j,l,m\}$. If we denote by $\uh_n$
the line inedexed by $n$ in $\H$, we get
\begin{eqnarray}
\Delta = \frac{3}{4}\sum_{i<j<l}\prob\{ i,j,l\in F\mbox{ and }
\uh_i=\uh_j = \uh_l\}\, .
\end{eqnarray}
The factor $3$ counts the number of different couples 
of pairs in $\{i,j,l\}$ and $1/4$ is the probability that
the corresponding entries in $\ub$ are different.
By computing the above probability and summing over 
$i,j,l$ we get $\Delta = K_2\gamma^3 L +O(1)$, thus proving 
Eq.~(\ref{eq:LowerBoundProb}).
%
%
\section{Polynomial transfer matrix algorithm}
\label{app:Algo}
\label{appendix_pol_alg}

Consider the constrained partition function
(\ref{eq:ConstrPartFun}) and the corresponding 
transfer matrix recursion (\ref{eq:RecPartFun}).
In this Appendix we shall consider only left-to-right 
iterations and drop the arrow $\rightarrow$ in subscripts.
We shall further set $n=2R$ and use the vector notation
$\vx_{j+1}^{j+n} = (x_{j+1},\dots,x_{j+n})$. 

The constrained partition function $Z_i(\vz)$ is just
the number of solutions in an inhomogeneous linear system,
obtained by retaining the lines of $\H$ with index
in $\{1,\dots,i-R\}$ (and the corresponding equations), and 
adding the $n$ equations $x_{i-n+1}=z_1$, \dots, $x_i = z_n$.
As a consequence, for all the choices of $\vz$ such that this
linear system has a solution, it has the same number of solutions as
corresponding homogeneous system. Further,
the number of solutions of the homogeneous system is a power
of $2$ (because it is the size of a linear space over $\Z_2$).
Finally, the vectors $\vz$ for which a solution exists form a linear space. 
Therefore, there exists a binary matrix $\A_i$ and an integer
$\Phi_i$ such that
\begin{eqnarray}
Z_i(\vz) = \left\{
\begin{array}{ll}
2^{\Phi_i} & \mbox{ if $\A_{i}\vz = \v0$,}\\
0 & \mbox{ otherwise.}
\end{array}\right.\label{eq:ConstrParametrization}
\end{eqnarray}
The matrix $\A_i$ can always be chosen as an $n\times n$ matrix by
eventually eliminating linearly dependent lines.

We therefore reduced the memory requirements from $\Theta(2^n)$
to $\Theta(n^2)$. We have now to show that the $\A_i$ and $\Phi_i$ 
can be computed recursively in polynomial time as well.
Consider the recursion (\ref{eq:RecPartFun}) and let 
$a_{i}=(a_{i,1},\dots, a_{i,n+1})$ be the binary vector defined
as follows. If $i-R+1\not\in F$ (the new equation added in the recursion
is not present), then $a_i \equiv \0t$. Otherwise,  
$a_{i,j}\equiv H_{i-R+1,i-2R+j}$ ($a_i$ encodes the newly added line 
of $\H$, properly shifted). Then define the $(n+1)\times (n+1)$
matrix $\B_i$ as follows
\begin{equation}
\B_i = \begin{pmatrix}      
 &  0 \\
\;\;\;\;\;\A_i\;\; & \vdots & \\
 &  0 \\
a_{i,1}  \cdots\cdots & a_{i,n+1}
\end{pmatrix} .
\end{equation}
Denote by $\b_i$ the first column of $\B_i$ (i.e. a column vector)
and by $\tB_i$ 
the $(n+1)\times n$ matrix formed by its last 
$n$ columns. By using Eq.~(\ref{eq:ConstrParametrization}) the 
recursion (\ref{eq:RecPartFun}) can be written as
\begin{eqnarray}
Z_{i+1}(\vz) = \,\, 2^{\Phi_i}\!\!\sum_{z_0\in\{\0t,\1t\}}
\ind\big(\b_i\,z_0+ \tB_i \,\vz = \0t\big)\, .\label{eq:NewRecursion}
\end{eqnarray}
Let us now consider two cases:
\begin{itemize}
\item If $\b_i = \0t$, then we get immediately the form 
(\ref{eq:ConstrParametrization}) for $Z_{i+1}$,
by letting $\Phi_{i+1}=\Phi_i+1$,
and $\A_{i+1}$ the matrix obtained by eliminating linear
dependencies among rows of $\tB_i$.
\item If $\b_i\neq \0t$, then there exists at least one vector
$\bh_i$ of dimension $(n+1)$ such that $\bh_i^T\b_i = \1t$ $\mod 2$.
The only non-vanishing term in the sum (\ref{eq:NewRecursion})
is therefore obtained for $z_0 = -\bh_i^T\tB_i\vz$ $\mod 2$. 
Substituting this value of $z_0$, we obtain
that $Z_{i+1}$ can again be written in the form 
(\ref{eq:ConstrParametrization}). 
The new matrix $\A_{i+1}$ is obtained by eliminating linearly dependent 
rows from $({\mathbf 1}-\b_i\bh_i^T)\tB_i$, while 
the number of solutions is updated by $\Phi_{i+1}=\Phi_i$.
\end{itemize}
In practice we found more convenient to reduce $\B_i$ in upper 
triangular form by gaussian elimination before 
computing $\A_{i+1}$ and $\Phi_{i+1}$ as just described. 

The initialization of the above recursion depends on the choice 
of boundary conditions. When using fixed boundary conditions with reference 
solution $\ux^{(0)} = \u0t$, we set
$\A_0 = {\mathbf 1}$ and $\Phi_0 = 0$. 

It is clear that the above procedure can be implemented in a time that is 
polynomial in the interaction range. Indeed the most complex operation
to be performed, consists in eliminating linearly dependent lines from the
matrix $\tB_i$, or $({\mathbf 1}-\b_i\bh_i^T)\tB_i$. This can be done
via gaussian elimination in time $O(R^3)$. The total complexity is therefore
$O(LR^3)$. 
%
%
\section{Analytical calculations}
\label{app:Analytical}

\subsection{Replicas}

In order to compute the ground state entropy and the point-to-set
correlation function, we shall follow the replica approach,
see \cite{MonassonReplicas}.
Each site $i\in\{1,\dots,L\}$ thus carries $n$ binary 
variables $\vx_i = (x^1_i,\dots,x_i^n)$ corresponding to the $n$
replicas. 

Let us consider first a particularly simple 
instance consisting of a single equation labeled by $i\in F$
and $2R+1$ variables on sites $j\in\{i-R,\dots,i+R\}$. Denote
by $\oc_i(\vx)$ the fraction of nodes $j$ such that
$\vx_j=\vx$. In formulae
\begin{eqnarray}
\oc_i(\vx) = \frac{1}{2R+1} \sum_{j=i-R}^{i+R}\ind(\vx_j=\vx)\, .
\end{eqnarray}
The probability that a randomly sampled equation at $i$ 
(with range $\{1-R,\dots,i+R\}$) is 
satisfied by all of the $n$ replicas, is a function of $\oc_i$, call it
$\F_{k,R}(\oc_i)$. For large $R$ it is easy to show that 
\begin{eqnarray}
\F_{k,R}(\oc) = \J_{k}(\oc) + \frac{1}{2R+1}\, \binom{k}{2}
[\J_{k}(\oc)-\J_{k-2}(\oc)] + O(R^{-2})\, ,
\end{eqnarray}
where
\begin{eqnarray}
\J_l(\oc) \equiv \sum_{\vx_1\dots\vx_l}\prod_{a=1}^n\ind(x^{a}_1\oplus\cdots
\oplus x^{a}_l=\0t)\; \oc(\vx_1)\cdots\oc(\vx_l)\, .
\end{eqnarray}

Consider now the full linear system and the partition function 
(\ref{eq:PartFun}). We shall implicitly assume periodic boundary conditions
in order to lighten the notations. Fixed boundary conditions can be recovered
by properly constraining the expressions that we will derive. It follows
from the above that 
\begin{eqnarray}
\E\{Z^n\} = \sum_{\{\vx_i\}} \prod_{i=1}^L[1-\gamma+\gamma \F_{k,R}(\oc_i)]\, .
\label{eq:Replicated}
\end{eqnarray}
Next we introduce two variables $\lambda_i(\vx)$, $c_i(\vx)$ 
indexed by $\vx\in\{\0t,\1t\}^n$ for each $i\in \{1,\dots,L\}$,
using the identity 
\begin{eqnarray}
1 = \int\! \de c_i(\vx) \int_{-i\infty}^{+i\infty}\frac{\de\lambda_i(\vx)}{2\pi i}
\, \exp\{-\lambda_i(\vx)(c_i(\vx)-\oc_i(\vx))\}\, .
\end{eqnarray}
This allows to perform the sum over $\vx_i$ in Eq.~(\ref{eq:Replicated}).
If we expand the resulting expression for large $R$ we get,
after some lengthy but straightforward calculations, 
\begin{eqnarray}
\E\{Z^n\} = \int\! \de c_i(\vx) \int_{-i\infty}^{+i\infty}\frac{\de\lambda_i(\vx)}{2\pi i}\;
\exp\Big\{-(2R+1)\, S_0[c,\lambda] -S_1[\,c\,] + O(1/R)\Big\}\, ,
\label{eq:PathIntegral}
\end{eqnarray}
where
\begin{eqnarray}
\!\!\!\!\!\!S_0[c,\lambda] & = & \frac{1}{2R+1}\sum_{i=1}^L
\left\{-\log[1-\gamma+\gamma\J_{k}(c_i)]+\sum_{\vx}\lambda_i(\vx)c_i(\vx)
-\log\left[\sum_{\vx}
e^{\sum_{j\in\D(i)}\frac{\lambda_j(\vx)}{2R+1}}
\right]\right\}\, ,\label{eq:S0}\\
\!\!\!\!\!\!S_1[\,c\,] & = & \frac{1}{2R+1}\sum_{i=1}^L
\gamma\binom{k}{2} \, \frac{\J_{k-2}(c_i)-\J_{k}(c_i)}
{1-\gamma+\gamma\J_k(c_i)}\, ,\label{eq:S1}
\end{eqnarray}
and we introduced the notation $\D(i) \equiv \{j:\, |i-j|\le R\}$.
%
%
\subsection{Mean field limit}

In the $R\to\infty$ limit, the integral (\ref{eq:PathIntegral})
is dominated by the saddle points of $S_0[c,\lambda]$.
We neglect for the moment the correction given by $S_1[c]$,
and look for a saddle point of the type
\begin{eqnarray}
c_i(\vx) = \varphi_i\, \delta_{\vx,\vx_0}+\frac{1}{2^n}(1-\varphi_i)\, ,
\;\;\;\;\;\;
\lambda_i(\vx) = \omega_i\, \delta_{\vx,\vx_0} + \frac{1}{2^n}\omega^0_i\, ,
\label{eq:Ansatz}
\end{eqnarray}
where $\vx_0 \equiv (\0t,\0t,\dots,\0t)$
and $\delta_{\vx,\vy}$ is the $n$-dimensional Kronecker delta function. 
There are several reasons for this Ansatz: $(i)$ The algebra
of functions of the form $f(\ux) = f_0+f_1\delta_{\vx,\vx_0}$
is closed; $(ii)$ This ansatz is known
to give the correct thermodynamic behavior for the
rXOR  ensemble (i.e. in the  mean-field limit); 
$(iii)$ Although it is replica symmetric,
it yields the correct one-step replica symmetry breaking
physics (it is a peculiarity of XORSAT that replica  symmetry can be
explicitely broken).

By substituting in Eq.~(\ref{eq:S0}) and letting $n\to 0$,
we get $S_0[c,\lambda] = A_0[\varphi,\omega]n\log 2 + O(n^2)$ 
where
\begin{eqnarray}
A_0[\varphi,\omega] = \frac{1}{2R+1} \sum_{i=1}^L
\left\{\gamma(1-\varphi_i^k)-\omega_i(1-\varphi_i)-e^{-\sum_{j\in\D(i)}
\frac{\omega_j}{2R+1}}\right\}\, .
\end{eqnarray}
By differentiating with respect to $\varphi_i$ and $\omega_i$ we
get the saddle point equations
\begin{eqnarray}
\varphi_i = 1-\frac{1}{2R+1}\sum_{j\in\D(i)}e^{-\sum_{l\in\D(j)}\frac{\omega_l}{2R+1}}\, ,\;\;\;\;\;\;\;\;\;\;
\omega_i = k\gamma\varphi_i^{k-1}\, .
\end{eqnarray}
The second of these equations can be used to eliminate $\omega_i$
from the action. 

If we finally assume that $\varphi_i$ only depends on $i$ on a scale of 
order $R$, we can set (with an abuse of notation) $\varphi_i = \varphi(i/R)$
and let $R\to\infty$ with $L=\ell R$, thus getting 
Eqs.~(\ref{eq:FSMeanField}) to (\ref{eq:FiniteRangeMFEq}).
%
%
\subsection{$1/R$ corrections}

In computing the $1/R$ corrections we shall assume the 
system to be homogeneous. For instance we can think of
imposing periodic boundary conditions, or  letting $L\to\infty$
at the outset. As a consequence, in the leading order calculation 
we have $\varphi_i = \varphi$ independent of $i$  and thus
$A_0[\varphi,\omega] = L\phi^{(0)}(\gamma;\varphi)$  with 
$\phi^{(0)}(\gamma;\varphi)$  as in Eq.~(\ref{eq:MFFreeEnergy}).
Hereafter $\varphi$ will denote a solution of the mean field equation
$\varphi = 1-\exp\{-k\gamma\varphi^{k-1}\}$, and we let
$\omega = k\gamma\varphi^{k-1}$.

There are two contribution to order $1/R$. The 
first one comes from the correction to the action and
is easy to compute. Substituting our Ansatz in
Eq.~(\ref{eq:S1}) and proceeding as in the previous 
Section we get $S_1[c] = A_1[\varphi]\,n\log 2+O(n^2)$,
where
\begin{eqnarray}
A_1[\varphi] = \frac{L}{2R+1}\gamma\binom{k}{2}\varphi^{k-1}(1-\varphi^2)\, .
\label{eq:ActionCorrection}
\end{eqnarray}

The second term comes from gaussian fluctuations 
around the saddle point. Let $c_i^*(\vx)$, $\lambda_i^*(\vx)$
denote the saddle point (\ref{eq:Ansatz}) and define
\begin{eqnarray}
c_i(\vx) = c_i^*(\vx) + \nu_i(\vx)\, ,\;\;\;\;
\lambda_i(\vx) = \lambda_i^*(\vx)+\xi_i(\vx)\, .
\end{eqnarray}
By expanding $S_0[c,\lambda]$ to second order around its saddle point,
we get
\begin{eqnarray}
S_0[c,\lambda] &= &S_0[c^*,\lambda^*]+\frac{1}{2(2R+1)}\sum_{i,j}
\sum_{\vx,\vy}\label{eq:QuadraticAction}\\
&&\Big\{A(\vx,\vy)\delta_{ij}\nu_i(\vx)\nu_j(\vy)
+\delta_{\vx,\vy}\delta_{i,j}
[\nu_i(\vx)\xi_i(\vy)+\xi_i(\vx)\nu_j(\vy)] -
B(\vx,\vy)\kappa_R(i-j)\xi_i(\vx)\xi_j(\vy)\Big\}\, ,\nonumber
\end{eqnarray}
where
\begin{eqnarray}
\kappa_R(i) =\left\{\begin{array}{ll}
(2R+1-|i|)/(2R+1)^2 & \mbox{ for $|i|\le 2R+1$,}\\
0 & \mbox{ otherwise.}
\end{array}
\right.
\end{eqnarray}
The coefficients appearing in Eq.~(\ref{eq:QuadraticAction})
have the form
\begin{eqnarray}
A(\vx,\vy) & = &A_1 + A_2\delta_{\vx,\vy}+A_3[\delta_{\vx,\vx_0}+
\delta_{\vy,\vx_0}]+A_4\delta_{\vx,\vx_0}\delta_{\vy,\vx_0}\, ,\\
B(\vx,\vy) & = &B_1 + B_2\delta_{\vx,\vy}+B_3[\delta_{\vx,\vx_0}+
\delta_{\vy,\vx_0}]+B_4\delta_{\vx,\vx_0}\delta_{\vy,\vx_0}\, ,
\end{eqnarray}
where (defining $z\equiv 1-\gamma(1-2^{-n})(1-\varphi^{k})$)
\begin{eqnarray}
A_1 & = & -\frac{1}{z}\, k(k-1)\gamma(1-\varphi^{k-2})+
\frac{1}{z^2}k^2\gamma^2 \,\frac{1}{2^n}(1-\varphi^{k-1})^2\, ,\\
A_2 & = & -\frac{1}{z}\, k(k-1)\gamma\varphi^{k-2}\, ,\\
A_3 & = & \frac{1}{z^2}\, k^2\gamma^2\, \frac{1}{2^n}(1-\varphi^{k-1})
\varphi^{k-1}\, ,\\
A_4 & = & \frac{1}{z^2} k^2\gamma^2\varphi^{2(k-1)}\, ,
\end{eqnarray}
and (defining $\Lambda_0 = (e^{\omega}-1+2^n)^{-1}$ and 
$\Lambda = 1-2^{n\Lambda_0}$)
\begin{eqnarray}
B_1 & = & -2^{n}\Lambda_0^2\, ,\;\;\;\;\;\;\;\;\; B_2 = \Lambda_0\, ,\\
B_3 & =  &-\Lambda\Lambda_0\, ,\;\;\;\;\;\;\;\;\;\;\; B_4 = \Lambda(1-\Lambda)\, .
\end{eqnarray}

The quadratic form in Eq.~(\ref{eq:QuadraticAction}) can be diagonalized both 
in position space (by Fourier transform) and in replica space (all the 
eigenvectors have the form $\zeta(\vx) = \zeta_0\delta_{\vx,\vx_0}+\zeta_1$).
One can therefore perform the gaussian integral, and let $n\to 0$. 
Putting this contribution together with the action correction,
cf. Eq.~(\ref{eq:ActionCorrection}), we finally get
the entropy correction
\begin{eqnarray}
\phi^{(1)}(\gamma) = -\binom{k}{2}\gamma\varphi^{k-2}(1-\varphi^2)-
\int^{+\infty}_{-\infty}\!\left\{
\log(1-a\, w(q))+\frac{b\,w(q)}{1-a\,w(q)}
\right\}\, \frac{\de q}{4\pi}\, ,
\end{eqnarray}
where 
\begin{eqnarray}
a & = &k(k-1) \gamma \varphi^{k-2}e^{-\omega}\, ,\\
b & = & -k(k-1)\gamma\varphi^{k-2}e^{-\omega}(1-e^{-\omega})+
k^2\gamma^2\varphi^{2(k-2)}e^{-\omega}\, .
\end{eqnarray}
%
%
\bibliographystyle{alpha}


\begin{thebibliography}{MPWZ02}

\bibitem{W1} T.~R.~Kirkpatrick and P.~G.~Wolynes, Phys.~Rev.~B~{\bf
36} (1987) 8552

\bibitem{W2}  T.~R.~Kirkpatrick, D.~Thirumalai, and P.~G.~Wolynes,
Phys.~Rev.~A~{\bf 40} (1989) 1045

\bibitem{DynamicsReview}J.-P.~Bouchaud, L.~F.~Cugliandolo, J.~Kurchan
and M.~M\'ezard, 
``Out of equilibrium dynamics in spin-glasses and other glassy system'', 
in {\it Spin Glasses and Random Fields}, A.~P.~Young
ed., (World Scientific, Singapore, 1997)

\bibitem{BiroliMonassonWeigt} Biroli G, Monasson R and Weigt M (2000)
    {\it Eur.~Phys.~J.~B} {\bf 14}, 551-568.

\bibitem{MarcGiorgioRiccardo} M\'ezard M, Parisi G and Zecchina R
  (2002) {\it Science} {\bf 297}, 812-815.

\bibitem{OursPNAS} F.~Krzakala, A.~Montanari, F.~Ricci-Tersenghi, 
G.~Semerjian, and L.~Zdeborova,
`Gibbs States and the Set of Solutions of Random Constraint Satisfaction 
Problems,' {\tt arXiv:cond-mat/0612365}, and Proc.~Natl.~Acad.~Sciences,
in press. 

\bibitem{W3} P.~G.~Wolynes, Jour.~Res.~NIST~{\bf 102} (1997) 187,

\bibitem{W4} X.~Xia, P.~G.~Wolynes, Proc.~Nat.~Acad.~Sci.~{\bf 97}, (2000) 2990

\bibitem{W5} X.~Xia, P.~G.~Wolynes, Phys.~Rev.~Lett~{\bf 86} (2001) 5526

\bibitem{BB2} J.-P.~Bouchaud and G.~Biroli,
J.~Chem.~Phys.~{\bf 121} (2004) 7347

\bibitem{MonSem} A.~Montanari and G.~Semerjian, 
J.~Stat.~Phys. {\bf 125}, 23 (2006).

\bibitem{FranzMontanari} S.~Franz and A.~Montanari, J.~Phys.~A {\bf 40}
(2007),  F251-F257

\bibitem{XOR_CS} N.~Creignou and H.~Daud\'e, Discrete~Appl.~Math.
{\bf 96-97} 41 (1999).

\bibitem{XOR} F. Ricci-Tersenghi, M. Weigt and R. Zecchina,
Phys.~Rev.~E {\bf 63}, 026702 (2001).

\bibitem{XOR_1} M. M\'ezard, F. Ricci-Tersenghi and R. Zecchina,
J.~Stat.~Phys. {\bf 111}, 505 (2003).

\bibitem{XOR_2}  S. Cocco, O. Dubois, J. Mandler and R. Monasson,
Phys.~Rev.~Lett. {\bf 90}, 047205 (2003).

\bibitem{NostroLettera} A.~Montanari and G.~Semerjian,
Phys.~Rev.~Lett.~{\bf 94}, 247201 (2005).

\bibitem{MontanariSemerjianBethe} A.~Montanari and G.~Semerjian,
J. Stat. Phys. 124, 103 (2006)

\bibitem{SchwarzMiddleton} J.~M.~Schwarz and A.~A.~Middleton,
Phys.~Rev. {\bf E 70} (2004) 035103 (R) 

\bibitem{FT1} S.~Franz and F.~L.~Toninelli, 
Phys.~Rev.~Lett. {\bf 92} (2004) 030602

\bibitem{FT2} S.~Franz and F.~L.~Toninelli,
J.~Phys.~A: Math. Gen. {\bf 37} (2004) 7433

\bibitem{FT3} S.~Franz and F.~L.~Toninelli, J.~Stat.~Mech. (2005) P01008

\bibitem{FranzParisi1d} S.~Franz and G.~Parisi,
Europhys.~Lett. {\bf 75}  (2006) 385-391

\bibitem{AlonSpencer}  N.~Alon and J.~Spencer, {\em The Probabilistic Method},
Wiley, New York, 1992.

\bibitem{MonassonReplicas} R.~Monasson, J.~Phys.~A
{\bf 31} (1998) 513-529


\end{thebibliography}

\newcommand{\etalchar}[1]{$^{#1}$}

\end{document}